\documentclass[aps,prl,reprint,twocolumn,noshowpacs,groupedaddress]{revtex4-1}  
\usepackage[]{graphicx}
\usepackage{floatrow}
\usepackage{amssymb}
\graphicspath{{../data/img/}}

\begin{document}
\title{Characteristic interfacial structure behind a rapidly moving contact line}
\author{Mengfei He and Sidney R. Nagel}
\affiliation{\textit{Department of Physics, the James Franck and Enrico Fermi Institutes, the University of Chicago, Illinois 60637, USA}}
\begin{abstract}
In forced wetting, a rapidly moving surface drags with it a thin layer of trailing fluid as it is plunged into a second fluid bath.  Using high-speed interferometry, we  find characteristic structure in the  thickness of this layer with multiple thin flat triangular structures separated by much thicker regions.  These features, depending on liquid viscosity and penetration velocity, are robust and occur in both wetting and de-wetting geometries.  Their presence clearly shows the importance of motion in the transverse direction.  We present a model using the assumption that the velocity profile is robust to thickness fluctuations that gives a good estimate of the thin gap thickness.  
\end{abstract}
\maketitle

\textit{Introduction:}  A solid entrains surrounding air along with its moving surface when it is pushed rapidly into a liquid bath.  In this process, known as ``forced wetting'', a three-phase contact line between the substrate, air and liquid is forced to move across the surface of the solid.  If the penetration velocity is high enough, the contact-line distorts downwards to create a pocket of air.  The effects of such entrainment are observed in the form of entrapped bubbles that can be problematic in printing and coating technologies.

\begin{figure}[h]
	\includegraphics[width=1\textwidth]{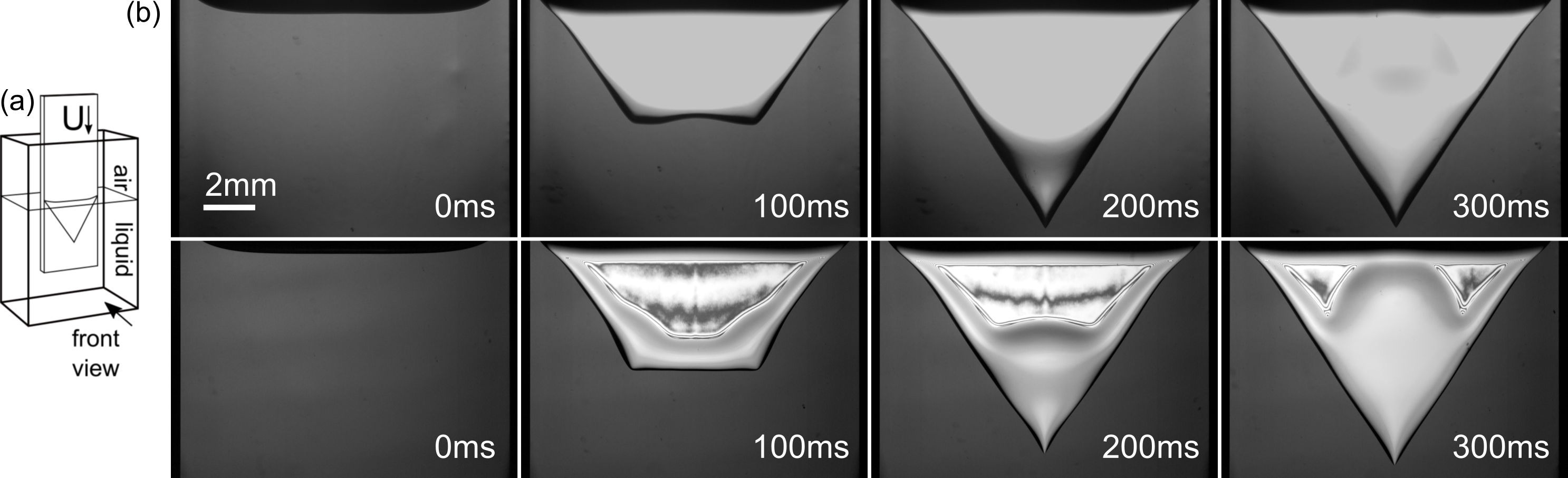}
	\caption{(a)  Schematic showing the ``V''-shaped steady-state contact line as viewed from the front.  (b) Images during the evolution of the ``V'' shape spaced $100 ms$ apart.  A $12.7mm$ wide tape travels vertically into a water/glycerol mixture of viscosity, $\eta = 226 cP$,  at $U=130mm/s$.  Top row:  Images using a white light.  Bottom row: Images using red light of coherence length $60\mu m$.  Interference patterns appear where the air pocket is thinner than $60 \mu m/2=30\mu m$.}
	\label{fig:formation}
\end{figure}

When the substrate velocity, $U$, is low, the contact line remains approximately level with the liquid surface.  At higher velocity, the line distorts and evolves towards a steady-state ``V'' shape~\cite{perry1967fluid,BURLEY1976901,blake1979maximum} shown schematically in Fig.~\ref{fig:formation}a.  The top row of Fig.~\ref{fig:formation}b shows images, spaced $100 ms$ apart, of the transient evolution to this shape.  The first frame shows the contact line immediately after a plastic substrate starts to move at fixed velocity into a liquid bath; the next images show the development towards the steady-state ``V'' shown in the last frame. 

These images, taken with white light, show the lateral evolution of the contact line but provide no information about the \textit{thickness} of the air gap at different points across its surface.  We obtain such information from interference fringes, which are visible when the optical path across the gap is less than the coherence length of the light.  In the bottom panel of Fig.~\ref{fig:formation}b, interference fringes appear for thicknesses less than $\approx 30 \mu m$.  These images reveal unexpected structure in the gap thickness that was not visible in the top panel.

As the contact line evolves, the air gap is thick near the edge and becomes thin and extremely flat in the center.  This flatness can be ascertained because over regions of approximately $5 mm$ in width there are only two fringes.  These correspond to equal-height contours, with a difference in thickness between successive bright fringes of $\approx 0.32 \mu m$.  Once the contact line has formed the ``V'' shape, the air gap continues to evolve until it reaches a steady shape shown in the last frame; at that point the air pocket has two very flat triangular structures that are symmetrically placed in the upper corners of the gap separated by an intervening thicker region. 

These features are very robust.  They appear regardless of the solid material (\textit{e.g.}, metal or plastic) and the fluid viscosity; they appear if the air is replaced by a second liquid.  More surprisingly, similar structures appear in de-wetting experiments where the liquid drains from the substrate as it is withdrawn from the bath.

\onecolumngrid

\twocolumngrid

We measured the dependence of the gap dimensions on the liquid viscosity, substrate width, and penetration velocity.  The \textit{absolute} thickness at different points in the gap were measured in order to characterize the three-dimensional structure of the air pocket.

\textit{Methods:}  In our experiments, we used flexible Mylar tape as the solid substrate.  The tape was held vertically as it was forced into (wetting) or pulled out of (de-wetting) the bath.  Vibrations and twist were minimized by supports located along the path of the tape.  These and the chamber walls were kept distant from the air pocket to avoid any interactions~\cite{vandre_carvalho_kumar_2012,vandre_carvalho_kumar_2014}.  Except where specifically stated otherwise, the tape width was  $12.7 mm$.  In each run, the tape velocity, $U$, was held constant at speeds between $50mm/s$ and $1000mm/s$.

The liquid bath consisted of water/glycerol mixtures whose viscosity, $\eta_{out}$, could be tuned between different runs by varying the relative concentration of the components: $26cP \le \eta_{out} \le 572cP$.  In order to check whether the structure of the gap was robust to the type of entrained fluid, we also replaced the air by a silicon oil of viscosity $0.65cP$.  Interfacial tensions, $\gamma$, and densities, $\rho$, were measured for different mixtures to be between $53mN/m$ and $66mN/m$ and between $1.21g/cm^{3}$ and $1.25g/cm^{3}$ respectively.

The absolute thickness, $H(x,y)$, of the air gap at different points on the surface $(x,y)$, was measured using high-speed interferometric imaging~\cite{PhysRevLett.107.154502}  from multiple wavelengths of light simultaneously~\cite{PhysRevLett.108.036101,PhysRevE.85.026315,PhysRevLett.109.264501}.  Once the thickness of the thin regions is known, the thickness of the gap in the thicker regions can be measured by counting fringes from a laser (see Supplemental Information).

\textit{Role of viscosity and evolution to steady state:}  The ``V'' shape of the steady-state contact line was quantitatively interpreted by Blake and Ruschak~\cite{blake1979maximum} in terms of a maximum contact-line velocity $U_{max}$ with which the liquid can wet the solid.  When $U>U_{max}$, the contact line is forced to tilt by an angle $\phi$ so that the normal velocity of the contact line does not surpass this threshold:
\begin{eqnarray}
	\label{eqn:umax}
	U cos\phi = U_{max}.
\end{eqnarray}
In the Supplemental information we compare $U_{max}$ to the \textit{normal relative velocity} at each point of the contact-line filmed during its evolution from an initial horizontal line into the final ``V'' shape. 

\begin{figure}[h]
	\includegraphics[width=1\textwidth]{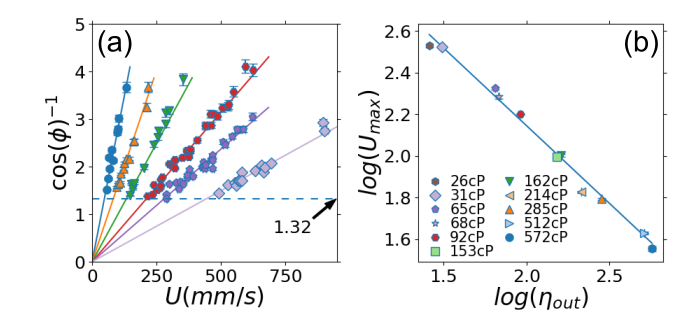}
	\caption{(a) $(cos\phi)^{-1}$ versus velocity $U$ for water/glycerol mixtures with viscosities between $26cP$ and $572cP$.  Solid lines: least-square-fits to Eq.~\ref{eqn:umax}.  (b) $U_{max}$ extracted from (a), versus $\eta_{out}$.  Solid line: $U_{max} \sim \eta_{out}^{-0.75}$.} 	
	\label{fig:angle}
\end{figure}

Figure~\ref{fig:angle}a shows $(cos \phi)^{-1}$ versus $U$ for liquids of different viscosities, $\eta_{out}$.  (Due to growing fluctuations in the contact line as $U$ decreases towards $U_{max}$, our data does not extend below $(cos\phi)^{-1} \approx 1.3$.) Figure~\ref{fig:angle}b shows that $U_{max}$ determined from Eq.~\ref{eqn:umax} varies as 

\begin{eqnarray}
	\label{eqn:umaxpowerlaw}
	U_{max} \sim \eta_{out}^{-0.75 \pm 0.03}.
\end{eqnarray}
This exponent is similar to that found in earlier works~\cite{perry1967fluid,WILKINSON19751227,BURLEY1976901,gutoff1982dynamic,BURLEY19841357,buonopane1986effect,Blake1997} but is larger than the value (between $1/3$ and $1/2$) suggested by Marchant et al.~\cite{PhysRevLett.108.204501}.

\begin{figure}[h]
	\includegraphics[width=1\textwidth]{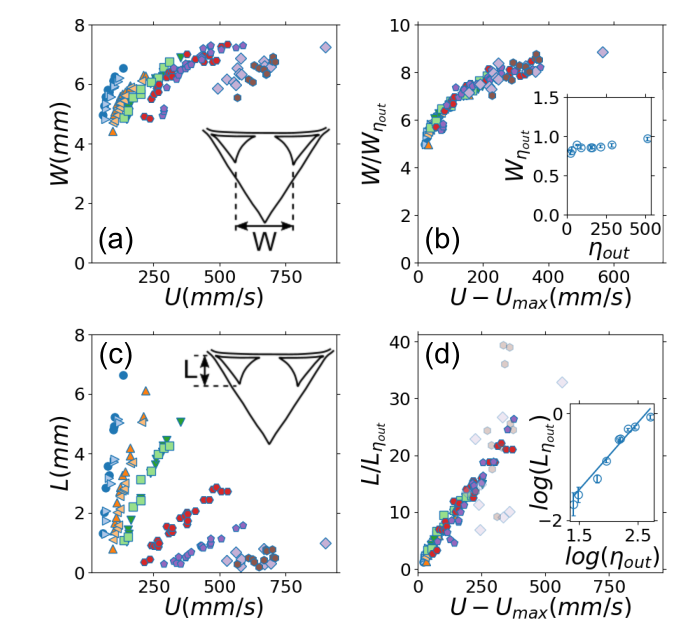}
	\caption{Lateral geometry of thin structures in the air gap.  (a) Distance between tips, $W$, versus $U$.  (b) Data collapse of the form $W/W_{\eta_{out}} = F_{W}(U-U_{max})$.  Inset: $W_{\eta_{out}}$ versus $\eta_{out}$.  (c) Vertical span $L$ versus $U$.  (d) Data collapse of the form $L/L_{\eta_{out}} = F_{L}(U-U_{max})$.  Inset: $L_{\eta_{out}}$ versus $\eta_{out}$.  Line $L_{\eta_{out}} \propto \eta_{out}^{1.3}$.}
	\label{fig:LWgeometry}
\end{figure}

\textit{Structure within the steady-state air gap:}  As the images in the bottom row of Fig.~\ref{fig:formation}(b) make clear, there is considerable structure in the thickness of the air gap, $H(x,y)$.  Most striking is the unexpected appearance of two flat steady-state triangular shapes in the upper corners of the last image.

Figure~\ref{fig:LWgeometry}a shows $W$, the tip separation of the triangular regions indicated in the inset, versus $U$.  Figure~\ref{fig:LWgeometry}b shows that the data can be collapsed to a form: $W/W_{\eta_{out}} = F_{W}(U-U_{max})$.  The inset shows the empirical fitting parameter $W_{\eta_{out}} \approx 1$ for all $\eta_{out}$.  $W$ appears to saturate at large $U$.
Figure~\ref{fig:LWgeometry}c shows $L$, the vertical span of the triangular regions, versus $U$.  The slope of $L$ versus $U$ increases with increasing bath viscosity.  In Fig.~\ref{fig:LWgeometry}d we collapsed the data using the form: $L/L_{\eta_{out}} = F_{L}(U-U_{max})$ where $L_{\eta_{out}}$ is an empirical fitting parameter shown in the inset.  With the exception of low $\eta_{out}$ where there are large fluctuations (shown as translucent), the collapse is good with $L_{\eta_{out}} \propto \eta_{out}^{1.3 \pm 0.1}$.

\begin{figure}[h]
	\includegraphics[width=1\textwidth]{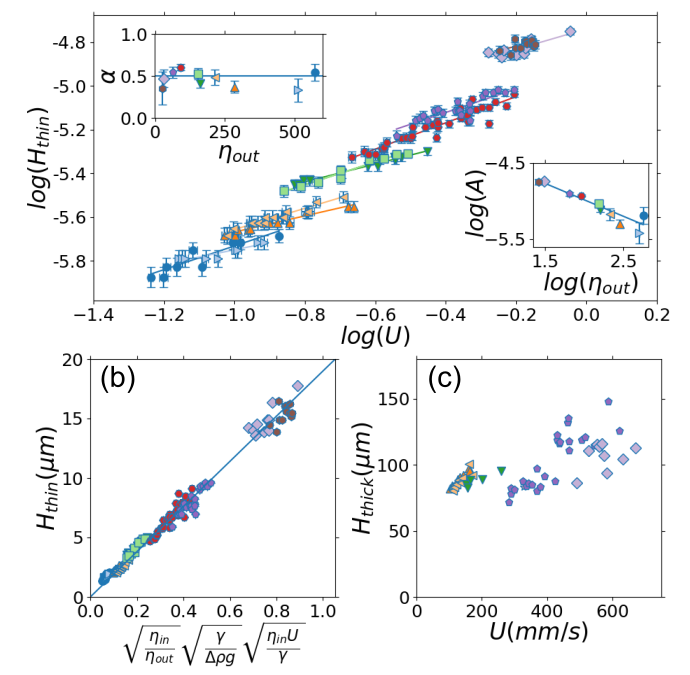}
	\caption{Thickness of air gap.  (a) Thickness of the thin triangular regions measured at their centers, $H_{thin}$, versus $U$.  Lines show fits for $ H_{thin} \propto A * U^{\alpha}$.  
Upper inset: $\alpha$ versus $\eta_{out}$.  The average $<\alpha> \approx 0.46 \pm 0.03$.  Lower inset: $A$ versus $\eta_{out}$.  Line shows fit $A \propto \eta_{out}^{-0.43}$.  (b) Data collapsed to Eq.~\ref{eqn:Hthinform}.  (c) Thickness at thickest point of the gap, $H_{thick}$, versus $U$.}
	\label{fig:thinthick}
\end{figure}

An average thickness of the air gap was previously estimated to be between $0.05\mu m$ and $0.9\mu m$ (\textit{e.g.}, see ~\cite{perry1967fluid,miyamoto1991mechanism}); in the case of a plunging liquid jet (instead of plunging solid) it was measured to be several microns~\cite{PhysRevLett.93.254501}.  No structure within the gap was reported.  

In Fig.~\ref{fig:thinthick}(a) we show $H_{thin}$, the absolute thickness of the gap in the center of the triangular regions obtained from the multi-wavelength interference method described in Supplemental Information, versus $U$.  These regions are extremely flat with a height variation of only $\Delta H_{thin} \approx 0.1\sim0.7 \mu m$ depending on the outer fluid viscosity.  As one might naively expect, $H_{thin}$ increases with increasing penetration velocity.  Each data set starts only when $U > U_{max}$ (from Fig.~\ref{fig:angle}(b)); thus as $\eta_{out}$ increases, the data range shifts to lower velocities.

The data for different values of $\eta_{out}$ \textit{do not} fall on top of one another but splay out and are roughly parallel to one another.  We fit each data set to the form: $H_{thin} = A*U^{\alpha}$.  The insets in Fig.~\ref{fig:thinthick}a show the least-square-fits of the parameters $\alpha$ and $A$ versus $\eta_{out}$.  The upper inset shows that the average $\alpha =0.46 \pm 0.03$.  The lower inset shows $A \propto \eta_{out}^\beta$ with a best-fit exponent $\beta = -0.43 \pm 0.07$ (solid line).  This suggests the form:  
\begin{eqnarray}
	\label{eqn:hthin}
	H_{thin} \propto \sqrt{\frac{\gamma}{\Delta\rho g}} \bigg(\frac{\eta_{in} U}{\gamma}\bigg)^{0.46 \pm 0.03} \bigg(\frac{\eta_{in}}{\eta_{out}}\bigg)^{0.43 \pm 0.07}
\end{eqnarray}
where$\Delta\rho g$ is the buoyancy force and $\eta_{in}$ is the \textit{inner} fluid viscosity (air in this case).

In order to understand this behavior, we model the air flow within the gap.  Because the contact line is stationary in the lab frame, the total flux of air must be zero; any air that is entrained by the substrate must have a return path to the surface.  (This is different from the situation of depositing a liquid layer on an infinite substrate pulled out of a bath~\cite{LANDAU1988141,derjaguin1943thickness}).  The case where there is no lateral flow so that the geometry is a two-dimensional wedge with fluid-substrate contact angle $\theta$ is treated by Huh \& Scriven~\cite{HUH197185}.  However, in the situation of forced wetting, with the ``V'' shape, there is clearly transverse flow.  The central, thickest part of the gap can accommodate the return of most of the entrained air so that in the very thin triangular regions there need not be any return flow.  In those thin regions, the entrained air can escape by flowing downwards towards the contact line and then sideways towards the central thicker part of the gap.  Our experiments show a different flow field than the scheme proposed by Severtson \& Aidun~\cite{severtson_aidun_1996}. 

We assume that the overall velocity of the liquid/liquid interface,  $U_{I}$, is still given by the two-dimensional results of Huh \& Scriven:
\begin{eqnarray}
	\label{eqn:UI}
	U_{I} = \zeta U \approx (1-D_{\theta}\frac{\eta_{in}}{\eta_{out}}) U
\end{eqnarray} 
where $D_{\theta}$ depends on contact angle $\theta$ and the second approximate expression is valid near $\theta \approx 2^{\circ}$ over the experimental range of $\eta_{in}/\eta_{out}$.  We assume that $U_{I}$ is determined by the average air flow in the gap, which is dominated by the thicker regions, and does not vary significantly across the surface.

In the thin regions, where the liquid interface is nearly vertical, the buoyancy force is balanced by the viscous forces in the inner fluid: $\eta_{in} \partial^{2} u(y) / \partial y^{2} = \Delta\rho g$ where $y$ is in the horizontal direction perpendicular to the substrate surface and the flow is in the $z$ direction.  Using the boundary conditions at the substrate $u(y=0) = U$ and at the liquid/liquid interface $u(y=H_{thin}) = U_{I}$ we find:
\begin{eqnarray}
	\label{eqn:poiseuille}
	&&u = \frac{\Delta\rho g}{2\eta_{in}}y^2+By+U \\
	\label{eqn:B}
	&&\text{with } B = -\bigg(\frac{(1-\zeta)U}{H_{thin}}+\frac{\Delta\rho g}{2\eta_{in}}H_{thin}\bigg).
\end{eqnarray}

Given an arbitrary $H_{thin}$ there is a solution satisfying both boundary conditions.  In order to determine which solution is chosen, we argue that the system selects the one that is invariant to thickness fluctuations.  Any noise or vibration in the flow can perturb $H_{thin}$ and disrupt the flows.  If the velocity profile is invariant to such fluctuations it will be stationary and robust against such noise with the least ``wandering'' of the system in the flow-field phase space.  Thus the system chooses the solution where $B$ is an extremum.  In that case, not only is $B$ independent of $H_{thin}$, but the profile has zero slope at the liquid/liquid interface.  Setting $dB/dH_{thin}=0$ gives:
\begin{eqnarray}
	\label{eqn:minimizeB}
	H_{thin} = \sqrt{2(1-\zeta)}\sqrt{\frac{\gamma}{\Delta\rho g}} \sqrt{\frac{\eta_{in} U}{\gamma}} 
\end{eqnarray}
(We note that this is the same solution as would be obtained by assuming minimum dissipation.)

Inserting Eq.~\ref{eqn:UI} for $\zeta$ leads to:
\begin{eqnarray}
	\label{eqn:Hthinform}
	H_{thin} = D_{\theta} \sqrt{\frac{\eta_{in}}{\eta_{out}}}\sqrt{\frac{\gamma}{\Delta\rho g}} \sqrt{\frac{\eta_{in} U}{\gamma}}. 
\end{eqnarray}
Comparing Eq.~\ref{eqn:Hthinform} to our data in Fig.~\ref{fig:thinthick}b shows an excellent agreement with $D_{\theta}\approx 19$ corresponding to $\theta = 2.7^{\circ}$.

\begin{figure}[h]
	\includegraphics[width=1\textwidth]{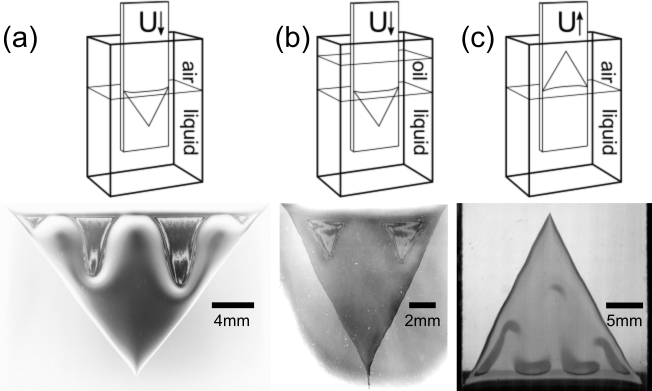}
	\caption{Robustness of structure for wetting and de-wetting geometries as shown in schematics.  (a) A wide $25.4mm$ tape moving from air into an $150cP$ water/glycerol bath showing four thin regions.  (b)  A $0.65cP$ silicone oil (replacing air) above a $60cP$ water/glycerol mixture.  A tape of width $12.7mm$ was used showing two thin triangular regions.  (c) A tape of width $25.4mm$ pulled out of a water bath into air showing three thin regions.  Gray scale inverted for clarity in (a) and (b).}
	\label{fig:robust}
\end{figure}

To see if $\theta = 2.7^{\circ}$ is reasonable for our experiment, we measure $H_{thick}$, the thickness of the air pocket at its maximum height (near the center of the ``V'' shape).  Figure~\ref{fig:thinthick}c shows that $H_{thick}$ is typically $\sim 100\mu m$, which is more than an order of magnitude larger than $H_{thin}$, and has large fluctuations.  From $H_{thick}$ and the dimensions of the ``V'' shape, we estimate $\theta$ 
to be between $1^{\circ}$ and $4^{\circ}$.  Thus, this model for the air flow in the thin regions is in quantitative agreement with our data. 
When the substrate width is varied, the number of thin regions in the air gap varies but leaves the distance, $W$, between them roughly constant.  Figure~\ref{fig:robust}a shows an image of such an entrained layer for a $25.4 mm$ wide tape (\textit{i.e.}, twice as wide as was used in the data shown above) with more thin-thick alternations across the tape surface.  Similar thin triangles appear if the air is replaced by another fluid as shown in Fig.~\ref{fig:robust}b where a $12.7 mm$ tape moves between a $0.65cP$ silicone oil and a $60cP$ water/glycerol mixture.  Two thin triangular regions appear in the upper corners of the ``V''.  
If we reverse the direction of $U$, so that the solid emerges from the bath and the liquid de-wets the substrate, 
a liquid film forms with three thin regions (now near the bottom) as shown in Fig.~\ref{fig:robust}c.  Rim-like structure behind the contact line in the longitudinal direction was previously seen in de-wetting~\cite{PhysRevLett.66.715,PhysRevLett.96.174504,PhysRevLett.100.244502,MALEKI2011359}, but no transverse thickness modulation was reported.  As with forced wetting, increasing the substrate width  produces more thin-thick alternations while leaving the distance $W$ between thin parts roughly constant.  Thus these thin triangular regions are a robust feature under both wetting and de-wetting conditions.

\textit{Summary:}  We have found an unexpected characteristic entrained layer in forced-wetting and de-wetting experiments.  This generic structure, consisting of extremely flat thin sections alternating with thick pockets, is stable and is controlled by viscosity contrast between the inner and outer fluids, the penetration velocity and width of the substrate.

Forced wetting is very different from wetting caused by a drop impacting a solid and spreading radially.  In that case, the contact line does not slide smoothly across the substrate; rather it moves by the nucleation of touch-down events near the encroaching interface.  The air gap breaks down at spots where the liquid film above it makes local contact with the surface~\cite{PhysRevLett.107.154502,PhysRevE.82.036302,thoroddsen2010bubble,kolinski2012}.  Our forced-wetting experiments show no such re-wetting holes.

For thin film problems such as gravitational flows, liquid films in rotating cylinders (\textit{i.e.}, the printer's instability), spinning drops and circular hydraulic jumps, the instability along the direction perpendicular to the general motion of the fluid has been observed and analyzed~\cite{huppert1982flow,0295-5075-10-1-005,Thoroddsen1997,rabaud1991wavelength,hosoi1999axial,PhysRevLett.63.1958,bush_aristoff_hosoi_2006}.  On the other hand, many attempts to understand wetting ignore motion transverse to the velocity of the substrate~\cite{RevModPhys.69.931}.  Such a simplification reduces the problem to a two-dimensional geometry.  While effective in describing the onset of the forced-wetting transition~\cite{snoeijer_andreotti_delon_fermigier_2007, delon_fermigier_snoeijer_andreotti_2008, PhysRevLett.93.094502, PhysRevLett.96.174504, PhysRevLett.100.244502,vandre2013mechanism,PhysRevLett.118.114502}, such analyses exclude the three-dimensional structures that emerge at later stages.  Our experiments show a pure two-dimensional analysis is no longer adequate in the steady state.  A uniform longitudinal velocity profile is unstable to transverse modulation; air in the thin regions of the gap flows in the direction of substrate motion and only in the thicker regions is there a return flow to the surface of the entrained air.  Our argument that assumes the velocity profile is robust to thickness fluctuations gives an excellent estimate for the flow profile in the thin regions.

\acknowledgments 
We thank Michelle Driscoll for early discussions of this work.  The work was primarily supported by the University of Chicago MRSEC, funded by the National Science Foundation under award number DMR-1420709 and by NSF Grant DMR-1404841.

\bibliographystyle{apsrev4-1}
\bibliography{references}

\end{document}


\title{Supplemental information}
\maketitle

In order for our experiments to run long enough to reach the steady-state regime, we used long spools of flexible plastic tape which was wound around a second reel, continuously driven by a motor (30 lb. in LEESON permanent magnet gearmotor).  We used commercial Mylar magnetic tapes of different widths (cassette tape for $6.4 mm$, VHS tape for $12.7 mm$ and recording tape for $25.4 mm$ widths) as the solid substrate.  For de-wetting, we also used Mylar sheets of $0.25mm$ thick cut to different widths.
 
The viscosity of the water/glycerol mixtures was measured by an Anton Paar MCR301 rheometer.  Interfacial tensions and liquid densities were measured by a KRUSS tensiometer.  The tape velocity, $U$, was measured using multiple images of the tape lit by a strobed light source.  The typical uncertainty was $2mm/s$.    

The liquid bath was held in an acrylic tank with a tilted front wall so that direct reflection from the wall would not interfere with the reflections from the sample under study.  All supports and walls were kept away ($\approx 30mm$ from the front, $\approx 5mm$ from the back, and $\approx 50mm$ from the two sides) from the entrained layer to minimize possible interactions from these boundaries~\cite{vandre_carvalho_kumar_2012,vandre_carvalho_kumar_2014}.  

\begin{figure}[h]
	\includegraphics[width=0.5\textwidth]{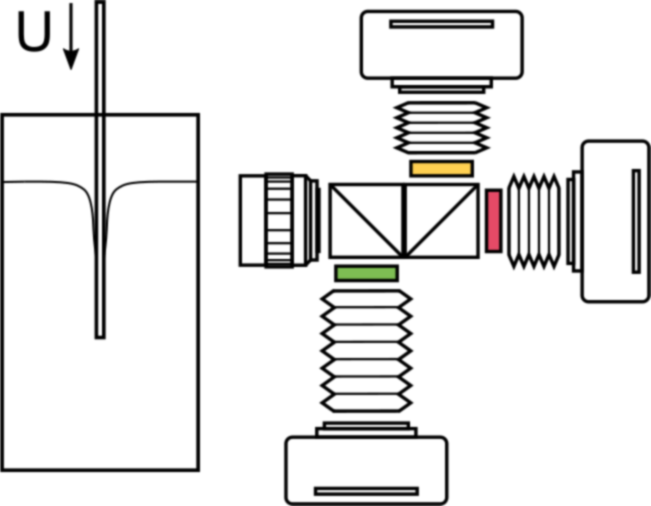}
	\caption{Interferometry setup for measuring absolute thickness.  Cameras (Nikon D7000) connected by beam splitters with one shared lens, were placed behind three narrowband filters with central wavelengths, $\lambda = 0.59 \mu m$ (amber), $0.63 \mu m$ (red) and $0.53 \mu m$ (green).  A fast, $100ns$, white-light pulse (palflash) illuminated the sample with the camera shutters open.}
	\label{fig:3cameras}
\end{figure}

The absolute thickness of the air gap as a function of position, $H(x,y)$, can be deduced without ambiguity while retaining the high precision of interferometry by using simultaneous information from three different wavelengths of light~\cite{PhysRevLett.108.036101,PhysRevE.85.026315,PhysRevLett.109.264501}.  Three cameras (Nikon D7000) connected by beam splitters were placed behind three narrowband filters with bandpass regions centered at wavelengths, $\lambda$ of $0.59 \mu m$ (amber), $0.63 \mu m$ (red) and $0.53 \mu m$ (green) respectively as shown in Fig.~\ref{fig:3cameras}.  Since distortion of images cause significant errors, we used a single lens (micro-NIKKOR 105mm f2.8) so that any distortion of the images would be nearly identical.  Prior to each measurement, particles of diameter $\approx 80\mu m$ were deposited onto the stationary tape surface to calibrate any remaining image distortion in the different cameras.  

The traveling substrate was illuminated from the front by a pulse of white light (palflash) for $\approx 100ns$ while the camera shutters were kept open.  The cameras would thus simultaneously capture three separate images at different wavelengths.  The interference patterns in these images go in and out of phase with one another as the gap thickness varies; there are periodic solutions where they are all in phase.  However, given the FWHM's of the filters, the contrast in the interference pattern of the associated camera's image provides an upper bound to the gap thickness of $\approx30\mu m$.  We chose the filters so that there is a unique solution, where all the interference fringes are in phase, that is less than this upper bound.  Using this information, the absolute gap thickness could be calculated.  We checked the accuracy of this method by measuring Newton's rings produced by a known lens.   

Synchronized with the white light pulse was a pulse from a green laser with $\lambda = 0.53 \mu m$.  Therefore in the green image, the interference patterns show up even in the thickest region of the gap.  Once a thickness value for a single spot was found, the rest of the shape could be determined by counting the green light fringes away from that point.  

\begin{figure}[H]
	\includegraphics[width=0.8\textwidth]{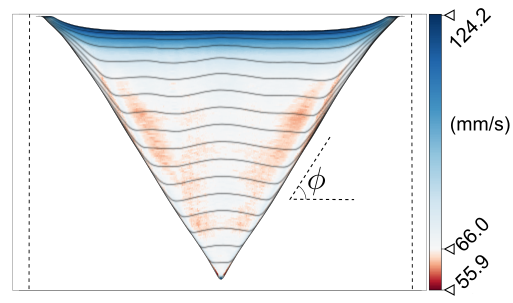}
	\caption{Normal relative velocity of the contact line with respect to moving substrate, calculated from a high-speed movie.  Solid lines: contact-line positions measured every $10ms$.  Dashed lines: substrate boundaries.}
	\label{fig:colormap}
\end{figure}

A Phantom VEO340 camera was used for high-speed movies to see how the structure formed and fluctuated.  A measured evolution of the contact line from an initial horizontal line into the final ``V'' shape is shown in Fig.~\ref{fig:colormap}($U=130mm/s$; liquid viscosity $=226cP$).  The solid lines are the contact-line positions measured every $10ms$.  From them the \textit{normal relative velocity} can be calculated.  Over most of the evolution the color is pale grey which, from the color map, indicates that the normal relative velocity $\approx U_{max}$($66mm/s$).  The blue colors show that while the contact line accelerates, the relative velocity initially exceeds $U_{max}$; red shows that $U < U_{max}$ where the contact line bends upwards.

\bibliographystyle{apsrev4-1}
\bibliography{references}